\documentclass[twocolumn,eqsecnum,preprintnumbers,nofootinbib]{revtex4-1}
\usepackage{amsmath,amssymb,amsfonts,times,graphicx,verbatim}

\newcommand{\beq}{\begin{equation}}
\newcommand{\eeq}{\end{equation}}
\def\bea{\begin{eqnarray}}
\def\eea{\end{eqnarray}}

\begin{document}

\title{Oscillating terms in the Renyi entropy of Fermi liquids}
\author{Brian Swingle}
\affiliation{Department of Physics, Harvard University, Cambridge MA 02138}
\author{Jeremy McMinis}
\author{Norm M. Tubman}
\affiliation{Department of Physics, University of Illinois at Urbana Champaign, Urbana IL 61820}

\date{\today}
\begin{abstract}
In this work we compute subleading oscillating terms in the Renyi entropy of Fermi gases and Fermi liquids corresponding to $2k_F$-like oscillations.  Our theoretical tools are the one dimensional formulation of Fermi liquid entanglement familiar from discussions of the logarithmic violation of the area law and quantum Monte Carlo calculations.  The main result is a formula for the oscillating term for any region geometry and a spherical Fermi surface.  We compare this term to numerical calculations of entanglement using the correlation function method and find excellent agreement.  We also compare with quantum Monte Carlo data on interacting Fermi liquids where we also find excellent agreement up to moderate interaction strengths.
\end{abstract}

\maketitle

\section{Introduction}
Fermi liquids are an extremely common form of quantum matter that appear in a wide range of physical systems including alkali metals, overdoped cuprate superconductors, and high density quark matter (before color superconductivity sets in).  Cousins of the Fermi liquid have been observed in the half filled Landau level \cite{hlr} and at a phenomenological level in layered organic salts \cite{2d_qsl_organics}.  The ubiquitous Fermi liquid has also played an important role in the recent exchange of ideas between quantum many-body physics and quantum information science.  This exchange has resulted in a deeper appreciation of the important role of entanglement, especially long-range entanglement, in the physics of quantum matter like Fermi liquids.  Long-range entanglement is important because it distinguishes interesting gapless phases like Fermi liquids and interesting gapped topological phases like fractional quantum Hall liquids from other more conventional phases like symmetry broken states that support long-
range classical correlation but not long-range entanglement.

Much of the discussion about entanglement has turned on a quantity known as the entanglement entropy.  We consider a large quantum system $AB$ divided into two components $A$ and $B$ (typically $A$ and $B$ are spatial regions, but other ``entanglement cuts" are possible).  The entanglement entropy is then the von Neumann entropy $S(A) = - \text{tr}_A(\rho_A \ln{(\rho_A)})$ of the state of $A$ where $\rho_A = \text{tr}_B(\rho_{AB})$.  When the state of the whole system, $\rho_{AB}$, is pure, the von Neumann entropy $S(A)$ measures the amount of entanglement between $A$ and $B$.  We are especially interested in the case when $\rho_{AB} = |G\rangle \langle G| $ with $|G \rangle$ the ground state of a local Hamiltonian.  It has also been profitable to consider the Renyi entropy,
\beq
S_n(A) = \frac{1}{1-n} \ln{\left(\text{tr}(\rho^n_A)\right)},
\eeq
which is actually a family of entropies labeled by $n$ and which gives complete information about the spectrum of $\rho_A$, the ``entanglement spectrum".

The basic rule governing the entanglement entropy in local ground states is the area law (for a review see Ref. \cite{arealaw1}).  For a system with short-ranged interactions in $d$ spatial dimensions, the area law states that the entanglement entropy of a region $A$ of linear size $L$ grows like $L^{d-1}$, that is like the area $|\partial A|$ of the boundary $\partial A$ of $A$.  Fermi liquids are extremely interesting from an entanglement perspective because they possesses long-range entanglement that manifests as a violation of the area law \cite{ee_f1,ee_f2,bgs_f1,bgs_f2,bgs_f3}.  Indeed, entanglement entropy in a Fermi liquid ground state scales like $L^{d-1} \ln{(L)}$ hence showing a logarithmic violation of the area law.  Gapless systems in one dimension, including Luttinger liquids and quantum critical points, also show a logarithmic violation of the area law \cite{geo_ent,eeqft}, while most other gapped and gapless systems in $d>1$ dimensions obey the area law.  In fact, the logarithmic violations
of the area law in $d=1$ gapless theories and Fermi liquids are intimately related as we review below \cite{bgs_f1}.

Another important development in the study of many-body entanglement has been the appearance of numerical studies of entanglement in a wide variety of systems.  Entanglement in one dimension has long been accessible using DMRG (for a review see Ref. \cite{dmrg_review}), and free fermions and bosons are accessible via the correlation matrix method \cite{CF-peschel}.  More recently, quantum Monte Carlo and tensor network calculations have permitted computations of entanglement in simple quantum magnets \cite{renyi_qmc_melko}, more complex spin liquid states\cite{qsl_ee_qmc}, topological states \cite{kagome_top_ee_dmrg}, and Fermi liquids \cite{qmc_renyi_FL}.  We are thus finally in a position to begin a substantive comparison between theory and (numerical) experiment for universal terms in the entanglement entropy.  We now have agreement between theory and experiment in one dimension and for certain simple topological phases in two dimensions, and various other predictions have been validated at a more
qualitative level, including a prediction of $L\ln{(L)}$ entropy in a spin liquid with spinon Fermi surface \cite{bgs_f1,qsl_ee_qmc} and an observation of corner terms and terms associated with symmetry breaking in quantum magnets (although the agreement here is not yet precise) \cite{anomalies_renyi_qmc,SB_ent_grover}.  Very recently the first calculations of Renyi entropy in interacting Fermi liquids were reported in Ref. \cite{qmc_renyi_FL}, with agreement at a quantitative level with previous theoretical predictions in Refs. \cite{bgs_f2,PhysRevX.2.011012} up to intermediate interaction strength.  Although it should be noted that discrepancies between these calculations and the theoretical predictions increase in the low density limit of the Fermi liquid, this represents one of the first examples in more than one dimension of precise quantitative agreement between theoretical and numerical computations of entanglement entropy in an interacting gapless system.

Previous work has focused on the leading logarithmic term in the Fermi liquid which was argued to depend only on the geometry of the interacting Fermi surface and on the geometry of $A$.  However, as may be expected on general grounds and as was evident in the data in Ref. \cite{qmc_renyi_FL} and elsewhere, there are also subleading oscillating terms in the Renyi entropy.  Here we compute these subleading terms in the Renyi entropy analytically for the free Fermi gas.  We also argue that the period of oscillation and exponent of the power law prefactor are unmodified by interactions.  We compare our results with extensive numerical data on free fermions and interacting Fermi liquids and find excellent agreement up to moderate interaction strengths.  Thus we establish in considerable detail a quantitative agreement between theory and numerics regarding entanglement in Fermi liquids, both for the leading logarithmic violation as well as for the subleading oscillating term.

This paper is organized as follows.  We first review the one dimensional aspects of Fermi liquid entanglement before deriving the form of the oscillating term in higher dimensions.  We give an alternate proof for our formula when the region geometry is a long strip and further elucidate the structure of the entanglement spectrum in this case.  Finally, we compare our theoretical predictions to numerical data for free and interacting Fermi liquids and find excellent agreement.

\section{Entanglement and the Fermi surface}

We now describe the theoretical framework for our calculations.  Let $R$ be the spatial region of interest in a Fermi liquid at zero temperature and let $\Gamma$ denote the interacting Fermi sea.  Once again, the Renyi entropy of region $R$ is defined as
\beq
S_n(R) = \frac{1}{1-n} \ln{\left(\text{tr}(\rho_R^n)\right)}.
\eeq
It is well known that the leading behavior of the Renyi entropy is $S_n \sim L^{d-1} \ln{L}$ where $d$ is the spatial dimension and $L$ is the linear size of $R$ \cite{ee_f1,ee_f2,bgs_f1,bgs_f2,bgs_f3}. Furthermore, the Widom formula,
\beq
S_n \sim \left(1 + \frac{1}{n}\right) \frac{1}{24} \int_{\partial R} \frac{1}{(2\pi)^{d-1}}\int_{\partial \Gamma} |n_x \cdot n_k| \ln{(L)},
\eeq
provides a precise characterization of the prefactor of the logarithmic term in terms of the geometry of the interacting Fermi surface $\partial \Gamma$ \cite{ee_f2,bgs_f1,bgs_f2}.  $n_x$ and $n_k$ are unit normals and the precise choice of length in the logarithm only modifies non-universal area law terms going like $L^{d-1}$.

This formula can be obtained by describing the Fermi surface as an infinite collection of one dimensional modes \cite{bgs_f1}.  Each such one dimensional mode is a gapless chiral fermion which contributes to the entanglement entropy like $ \frac{\ln{(\ell)}}{6}$ where $\ell$ is some effective length (see below).  Adding up these contributions for each point on the real space boundary $\partial R$ and each mode on the Fermi surface $\partial \Gamma$ leads to the Widom formula above.  Even the dependence on the Renyi parameter $n$ is predicted by the theory since the Renyi entropy of a single interval in a CFT is $\frac{n+1}{2n} \frac{\ln{(\ell)}}{6}$ \cite{ee_cft}.  Note also that each mode experiences many different effective lengths corresponding to different one dimensional cuts through the real space region, but to logarithmic accuracy we may replace all such lengths with any particular representative of the linear dimension $L$.

It is also known that the Renyi entropy has a subleading oscillating term in one dimensional Luttinger liquids.  This term is analogous to Friedel oscillations and hence occurs at momentum $2 k_F$ \cite{ee_osc_1d}.  For free fermions this term has the form
\beq \label{renyi1dosc}
S_n^{d=1} \sim f_n \frac{\cos{(2 k_F \ell)}}{(2 k_F \ell)^{\beta_n}}
\eeq
with $\beta_n = 2/n$ and
\beq
f_n = \frac{2}{1-n} \left(\frac{\Gamma((1+n^{-1})/2)}{\Gamma((1-n^{-1})/2)}\right)^2
\eeq
($\Gamma(z)$ is here the gamma function) \cite{ee_osc_1d}.  Below we will use our one dimensional formulation of higher dimensional Fermi liquid entanglement to demonstrate the existence of similar oscillating terms in the entanglement entropy in higher dimensions.

First, however, we give a quick derivation of the 1d result using conformal field theory.  The 1d electron operator may be decomposed as $c(x) = c_L (x) e^{-i k_F x} + c_R(x) e^{i k_F x}$ where $c_{L/R}$ are the left and right moving halves a free relativistic fermion. They are the slowing varying fields entering the low energy description which we can use to compute universal terms in the entanglement entropy.  The typical way we proceed is to introduce twist fields or otherwise study the system on an n-sheeted surface to compute $\text{tr}(\rho^n)$.  Here we simply note that an important feature of this setup is that translation symmetry is broken for $n\neq 1$.  Furthermore, the singular branch points can in general produce localized \textit{relevant} perturbations bound to them.  These perturbations produce corrections to the leading CFT scaling of Renyi entropy (there are also bulk irrelevant operators that contribute) since they change the free energy in the branched
background.

For example, the action is perturbed to $\mathcal{S} = \mathcal{S}_{CFT} + g \phi(z_1) + g \phi(z_2)$ where $z_{1,2}$ are two branch points.  The path integral is \bea
&& Z = e^{-F} = \int D(\text{fields}) e^{- \mathcal{S[\text{fields}]}} \cr \nonumber \\
&& = Z_{CFT} (1 - g \langle (\phi(z_1)+ \phi(z_2))\rangle + \cr \nonumber \\
&& \frac{g^2}{2} \langle (\phi(z_1)+ \phi(z_2))^2\rangle + ... ) .
\eea
Note that we have not been careful about the $n$ dependence in this schematic expression.  Since $\langle \phi \rangle =0$ the first correction to the free energy (upon re-exponentiating the series and with the usual story about disconnected diagrams) is of order $g^2$ and is essentially a correlator of the form $\langle \phi(z_1) \phi(z_2)\rangle$.  To be specific, this is the part which depends on the separation of the branch points.  We now go through this calculation in more detail for two branch points.

Consider a finite interval of length $\ell$ in the free fermion CFT.  To produce an n-branched surface with branch points at $z_1$ and $z_2$ with $z_1 - z_2 =\ell$, we can consider the conformal transformation given by
\beq \label{branch}
w = \left(\frac{z-z_1}{z-z_2}\right)^{1/n}
\eeq
where $w$ is a coordinate on a plane while $z$ is the branched coordinate.  Indeed, we see that winding $z$ around $z_1$ or $z_2$ winds $w$ by a phase of $2\pi/n$, and hence we must wind $z$ by $2\pi n$ to wind $w$ by $2 \pi$.  We now wish to compute correlations of operators inserted at $z_1$ and $z_2$ (the induced defect operators) to find corrections to the entropy.  Given a primary field $\phi$ of dimension $\Delta$ we wish to find
\beq
\langle \phi(z) \phi(z') \rangle = \left|\frac{dw}{dz}\right|^\Delta \left|\frac{dw}{dz}\right|^\Delta \langle \phi(w) \phi(w')\rangle
\eeq
where the equality follows from conformal invariance under the transformation in Eq. \ref{branch}.

This equation is valid for all $z$ and $z'$ but we specifically want $z=z_1 +\epsilon$ and $z'=z_2+\epsilon$ with $\epsilon$ a UV regulator.  Assuming $\ell \gg \epsilon$ we obtain the following formulas:
\beq
\frac{dw}{dz} = \frac{1}{n} \left(\frac{\epsilon}{\ell}\right)^{1/n} \frac{1}{\epsilon},
\eeq
\beq
\frac{dw'}{dz'} = \frac{1}{n} \left(\frac{\ell}{\epsilon}\right)^{1/n} \frac{-1}{\epsilon},
\eeq
and
\beq
\langle \phi(w) \phi(w')\rangle = \frac{1}{|w-w'|^{2\Delta}} = \left(\frac{\epsilon}{\ell}\right)^{2\Delta/n}.
\eeq
Putting everything together we find
\beq
\langle \phi(z_1) \phi(z_2) \rangle = \frac{1}{(n\epsilon)^{2 \Delta}} \left(\frac{\epsilon}{\ell}\right)^{2\Delta/n}
\eeq
which explicitly shows a correction of the form $\ell^{-2 \Delta /n}$.  Finally, to make contact with the Fermi gas result we must identify the relevant operator, but this operator is just the $2 k_F$ density operator given by $e^{2 i k_F x} c^\dagger_L(x) c_R(x)$ coming from the expansion of $c^\dagger (x) c(x)$.  This operator has dimension $\Delta = 1$ and has an oscillating phase that is explicitly displayed.  Thus the oscillating term term in the entanglement entropy is indeed interpretable as a kind of $2 k_F$ density response, albeit in a branched space (which accounts for the strange scaling dimension).  Obtaining the prefactor for a given model requires more work since the field theory gives a cutoff dependent answer for the prefactor (which is anyway non-universal), but see Ref. \cite{ee_osc_1d} for a calculation for free fermions.

We now return to the extension to higher dimensions.  For concreteness, we set $d=2$ and take $R$ to a disk of radius $L$ and $\Gamma$ to be a disk of radius $k_F$.  The extension of our results to arbitrary region shape (the analog of the Widom formula) is completely straightforward and will be recorded later.  First, we must be more careful about the effective length $\ell$ since we are studying subleading terms.  The appropriate choice is $\ell = \ell(x,k) = 2 L |\cos{\theta}|$, where $\theta$ is the angle between $n_x$ and $n_k$, which is nothing but the chordal distance across the circle (parallel to $n_k$) at angle $\theta$ \cite{bgs_f2,bgs_f3}.  This choice is the effective one dimensional distance experienced by a mode propagating in the $n_k$ direction starting at angle $\theta$ on the circle and reproduces the correct thermal entropy to entanglement entropy crossover function \cite{bgs_f2,bgs_f3}.

The subleading oscillating term is then given by
\beq
f_n \frac{1}{2 \pi} \frac{1}{4} \int_{\partial R} \int_{\partial \Gamma} |n_x \cdot n_k| \frac{\cos{(2 k_F \ell(x,k))}}{(2 k_F \ell(x,k))^{\beta_n}}.
\eeq
Plugging in our various expressions we find
\bea
&& S_n \sim f_n \frac{4 k_F L}{8} \frac{1}{(4 k_F L)^{2/n}} \cr \nonumber \\
&& \times \int_{-\pi/2}^{\pi/2} d\theta (\cos{\theta})^{1-2/n} \cos{(4 k_F L \cos{\theta})}.
\eea
The final integral may be simplified when $4 k_F L$ is large since the integrand is rapidly oscillating.  Focusing on the part of the integral near $\theta=0$ we may write
\beq
\int d\theta e^{i u \cos{\theta}} \sim \int d\theta e^{i u (1- \theta^2/2)} \sim \sqrt{\frac{2\pi}{u}} e^{i u + i \phi}
\eeq
where $\phi = \pm \pi/4$ is an unimportant phase.  Note that this leading estimate is completely independent of $n$ as regards the integral over $\theta$.  Using this formula in our main expression gives, for $4 k_F L$ large, the result
\beq
S_n \sim \frac{\sqrt{2 \pi} f_n}{8} (4 k_F L)^{1/2 - 2/n} \cos{(4 k_F L + \phi')}
\eeq
where $\phi'$ is another phase.  Thus we have a prediction for the prefactor, power lay decay, and period of oscillation of the oscillating term in the Renyi entropy for a free Fermi gas.

\section{Interactions and general region shapes}

When interactions are turned on we expect that the emergent $U(1)^\infty$ symmetry (the fact that the Fermi liquid fixed point is essentially free) will protect the exponent of the power law, but the numerical prefactor may be modified \cite{bgs_f2,bgs_f3}.  For example, in one dimension the prefactor depends on the Luttinger parameter \cite{ee_osc_1d}, but while the power law decay is also modified in one dimension, in higher dimensions the power law should remain unchanged since quasiparticles remain sharp.  The momentum of the oscillation will remain at the interacting $2 k_F$, that is, while interactions may change the non-interacting Fermi surface, the correct momentum is always $2 k_F$ for the physical Fermi surface.

The $U(1)^\infty$ symmetry refers to the fact that Landau's Hamiltonian for the Fermi liquid takes the form
\beq \label{forwardH}
H_{FL} = \sum_k (\epsilon_k -\mu) n_k + \frac{1}{2} \sum_{k k'} f_{kk'} n_k n_{k'}
\eeq
which has $[H_{FL},n_k] = 0$ for all $k$ and hence an infinite number of conserved quantities.  Indeed, the ground state of this interacting Hamiltonian is a free fermion wavefunction with a Fermi sea $\Gamma$ obtained by solving the self consistent equations
\beq
\tilde{\epsilon}_k \leq 0, k \in \Gamma
\eeq
and
\beq
\tilde{\epsilon}(k) = \epsilon_k - \mu + \sum_{k'\in \Gamma} f_{kk'}.
\eeq
For example, if we work in a rotationally invariant system and impose Luttinger's theorem (by adjusting the chemical potential) then the interacting Fermi surface always coincides with the free surface which is in turn determined solely by the density.

More generally, although the Fermi surface can change as interactions are tuned (e.g. in a lattice model with non-spherical Fermi surface), the entanglement entropy always obeys the Widom formula evaluated on the physical interacting Fermi surface.  Similarly, all subleading corrections to the entropy in the toy model in Eq. \ref{forwardH} are those of a Fermi gas with Fermi surface $\Gamma$.  However, we emphasize that this is not necessarily the correct answer for a physical Fermi liquid since the pure forward scattering model in Eq. \ref{forwardH} requires long-range interactions and hence is not in the same universality class as a short-ranged Fermi liquid.  This is reflected physically in the fact that while Eq. \ref{forwardH} has an exact $U(1)^\infty$ symmetry, the same symmetry in a short-ranged Fermi liquid is only emergent at low energies.

Our argument for universality has recently been explicitly validated in Ref. \cite{entsumrule}.  That work examined a model of a Fermi liquid in which the quasiparticle residue could be continuously tuned to zero at a quantum critical point, but where density and current correlation functions were those of a free Fermi gas.  In the language of Fermi liquid theory, the Landau parameters have been tuned to zero, but the quasiparticle residue is non-trivial.  Nevertheless, the system is a non-trivial interacting Fermi liquid and Ref. \cite{entsumrule} showed that the Widom formula for the leading logarithmic violation is still exactly obeyed.  Furthermore, the oscillating terms we considered here are also present, unmodified, in the particular realization of an interacting Fermi liquid in Ref. \cite{entsumrule}.  This suggests that the prefactor of the oscillating term can be expected to be near the free result provided the Landau parameters are small even if the quasiparticle residue is tiny.

Now we turn to the general formula for the subleading oscillating term in the Renyi entropy of a free Fermi gas for a convex but otherwise arbitrary region shape.  Again, $R$ is the real space region of linear size $L$ and $\Gamma$ is the spherical Fermi sea.  We can also generalize to more complex Fermi seas in a straightforward way.  Let $x$ and $k$ denote points on $\partial R$ and $\partial \Gamma$ respectively, and define the effective length $\ell(x,k)$ to be the length of the line segment given by the intersection of the line $\{x+n_k s| s\in (-\infty,\infty)\}$ and $R$.  The convexity of $R$ guarantees that this intersection is a single line segment.  It can also be checked that this definition reduces to our prescription for $\ell$ for a spherical region $R$ given above. The subleading correction to the Renyi entropy is then
\beq
S_n \sim \frac{f_n}{2 \pi} \frac{1}{4} \int_{\partial R} \int_{\partial \Gamma} |n_x \cdot n_k| \frac{\cos{(2 k_F \ell(x,k))}}{(2 k_F \ell(x,k))^{2/n}}.
\eeq

We have already applied this formula to case when $R$ is a disk.  It is also enlightening to consider a long strip region.  Thus suppose the region $R$ is a long strip of length $L$ and width $W$ with $L \gg W$, and let $\theta$ be the angle between the Fermi surface normal and real space normal.  The effective distance is found to be $\ell = W/|\cos{\theta}|$ so long as $\ell \ll L$ which we take to be essentially infinite.  Alternatively, we can consider a strip that wraps completely around the cycle of a torus of length $L$ in which case translation invariance in the $L$ direction is manifestly preserved.  Within our approximation we find
\beq
S_n \sim \frac{f_n k_F L}{2\pi} \int_{-\pi/2}^{\pi/2} d\theta \cos{\theta} \left(\frac{\cos{\theta}}{2 k_F W}\right)^{2/n} \cos{\left(\frac{2 k_F W}{\cos{\theta}}\right)}.
\eeq
Assuming $2 k_F W \gg 1$ we may perform the $\theta$ integral as above by focusing on the region near $\theta = 0$.  The result is
\beq
S_n \sim \frac{f_n k_F L}{2\pi} \left(\frac{1}{2 k_F W}\right)^{2/n}\sqrt{\frac{2\pi}{2k_F W}}\cos{(2k_F W)}.\label{theta_int_1}
\eeq

Part of the reason why the strip case is interesting is that we may obtain the above result in another way.  We can also say a great deal about the entanglement spectrum of the strip.  The results are, however, restricted to free theories only.  Thus consider again a free Fermi gas with spherical Fermi surface and examine the fermion two-point function $G(x-y) = \langle c^\dagger(x) c(y)\rangle$.  This function may be obtained from a Fourier transform of the occupation number $n_k$ as
\beq
G = \int \frac{d^2 k}{(2\pi)^2} n_k e^{i k \cdot (x-y)}
\eeq
where $n_k = \theta(k_F - |k|)$.
Our interest in this function is that, by virtue of Wick's theorem, it completely determines the reduced density matrix of a region $R$ provided we restrict $x,y\in R$.  This is the standard correlation matrix method (see the appendix).

Let $\hat{K}$ be the entanglement Hamiltonian ($\rho = e^{-\hat{K}}$) for the infinite strip (length $L$, see above).  We know that $\hat{K}$ has the form
\beq
\hat{K} = \sum_{x,y \in R} K(x,y) c^\dagger(x) c(y)
\eeq
because of Wick's theorem.  Furthermore, the ``matrix" $K$ is related to $G$ via $G = \frac{1}{e^K+1}$ (proven by diagonalizing the ``matrix" $G$ with $x,y$ restricted to $R$).  Translation invariance in the $L$ direction, call it the $2$ direction, enables us to write
\beq
K = K(x^1,y^1,x^2,y^2) = K(x^1,y^1,x^2-y^2,0)
\eeq
or upon Fourier transforming over $x^2-y^2$
\beq
\hat{K} = \sum_{0 \leq x^1,y^1 \leq W, k} K_{1d}(x^1,y^1,k) c^\dagger(x^1,k) c(y^1,k).
\eeq

The new matrix $K_{1d}$ is related to the partial Fourier transform $G$.  Consider the mixed position/momentum basis function
\bea
&& G_{1d}(x^1-y^1,k) = \sum_{x^2} G(x^1-y^1,x^2) e^{-i k x^2} \cr \nonumber \\
&& = \int_{-\sqrt{k_F^2-k^2}}^{\sqrt{k_F^2-k^2}} \frac{dq}{2\pi} \theta(k_F - \sqrt{q^2+k^2}) e^{i q(x^1-y^1)}
\eea
which is nothing but the two-point function of a \textit{one dimensional} Fermi gas with Fermi momentum $k_F^{1d}(k) = \sqrt{k_F^2-k^2}$.  Because both $K$ and $G$ are partially diagonalized by the Fourier transformation in the $2$ direction we immediately have that
\beq
G_{1d}(k) = \frac{1}{e^{K_{1d}(k)}+1}.
\eeq
Thus $K_{1d}$ is the single particle entanglement Hamiltonian for a one dimensional Fermi gas of Fermi momentum $k_F^{1d} = \sqrt{k_F^2 - k^2}$ on an interval of length $W$.  The full entanglement spectrum of the two dimensional strip is also now known in terms of the one dimensional spectrum e.g. we have a complete one dimensional spectrum for an interval of length $W$ for each value of $k \in [-k_F, k_F]$.

Using this information we can immediately check the leading $L \log{(W)}$ term in the entropy.  The entropy from each value of $k$ is $\ln{(W)}/3$ and hence the total entropy is (to leading order)
\beq
S =  L \int_{-k_F}^{k_F} \frac{dk}{2\pi} \frac{\log{(W)}}{3} = \frac{k_F L}{3\pi} \ln{(W)}.
\eeq
A quick calculation with the Widom formula gives
\bea
&& S = \frac{1}{12}\frac{1}{2\pi} (2 \pi L) (2 k_F) \int_{-\pi/2}^{\pi/2} d \theta \cos{\theta} \ln{(W)} \cr \nonumber \\
&& = \frac{k_F L}{3 \pi} \ln{(W)}.
\eea

Returning now to the oscillating term, we can use Eq. \ref{renyi1dosc} to estimate the oscillating term for the strip in a different way.  Each one dimensional spectrum identified above will contribute an oscillating term but with a variable $k_F^{1d}$.  The oscillating term is thus
\beq
S_n \sim f_n L \int_{-k_F}^{k_F} \frac{dk}{2\pi} \frac{\cos{(2 \sqrt{k_F^2-k^2} W)}}{(2 \sqrt{k_F^2-k^2} W)^{\beta_n}}.\label{theta_int_2}
\eeq
As we have now repeatedly observed, if $k_F W$ is large then the integrand is rapidly oscillating and the integral is dominated by values of $k$ near zero.  Performing the effective Gaussian integral over $k$ we find
\beq
S_n \sim \frac{f_n L}{2\pi} \sqrt{\frac{\pi k_F}{W}} \frac{\cos{(2k_F W)}}{(2 k_F W)^{2/n}}
\eeq
or
\beq
S_n \sim \frac{f_n k_F L}{2\pi} \sqrt{\frac{2 \pi}{2 k_F W}} \frac{\cos{(2k_F W)}}{(2 k_F W)^{2/n}}
\eeq
which is identical to our previous result.  Note how the two methods obtain the same final form by integrating over either an effective length or an effective Fermi momentum.

\section{Comparison between theory and numerics}

We will now compare our theoretical predictions with numerical data for the Renyi entropy of the free Fermi gas for two 2D geometries, strips and circles. All numerical data was produced using the correlation function technique extended to continuum Hamiltonians, and fits to the data were performed using a standard non-linear least squares (NLLS) package \cite{CF-peschel,Renyi-mol,Gnuplot}. The data is presented in two plots with the corresponding parameters from the fits provided in the tables. Computational details for computing Renyi entropies and the fitting procedure can be found in Appendix \ref{comp-detail}.


For the circular geometry, we work in units where $k_F = 2$ so that the density of a spin polarized gas is $n = k_F^2/(4 \pi) = 1/\pi$.  This implies that the average number of particles in the real space circle is $\langle N \rangle = \pi L^2 n = L^2$.  In the figure we show Renyi entropies of a free Fermi gas computed numerically for disks with up to $L=6$.  The numerical data is fit to a two parameter function having the form
\beq
S_n(L) = a_1 L \ln{(L)} + a_2 L + a_3 \frac{\cos{(a_4L+a_5)}}{L^{2/n-1/2}},\label{sph-param}
\eeq
with known parameters,
\beq
a_1 = \frac{n+1}{3n}
\eeq
(from the Widom fomula),
\beq
|a_3| = |f_n|\frac{\sqrt{2\pi}}{8^{1/2+2/n}}
\eeq
and $a_4=8$. The parameters $a_2$ and $a_5$ are not predicted by our theory and are determined by fitting the numerical data.  This form of the scaling law includes our new prediction for the subleading oscillations as well as the leading order term from the Widom conjecture and the area law term.
In Table \ref{sph-fit} we provide the parameters determined by our analysis as well as those from the fit. Root mean square residual (RMSR) values, the square root of the sum of the residuals divided by the number of degrees of freedom in the fit, are provided as a metric for fit quality.
Although we might expect other subleading terms not included in our theory to cause noticeable deviations, especially in the limit of small L, the fits remain close to the data over the entire range, as demonstrated in Figure \ref{sph-fit}.

\begin{table}
\begin{center}
\begin{tabular}{|c|c|c|c|c|c|c|}
\hline
$n $ & $a_1$ & $(a_2)_{fit}$ & $a_3$ &  $a_4$ & $(a_5)_{fit}$ & $RMSR$ \\ \hline
1 & 0.667 & 1.774 &  0         &  0  & 0        & 0.011\\
2 & 0.500 & 1.384  &  0.0253    &  8  & 0.60 & 0.008\\
3 & 0.444 & 1.242  &  0.0566    &  8  & 0.65  & 0.015\\
4 & 0.4167& 1.171  &  0.0765    &  8  & 0.67  & 0.027\\
5 & 0.400 & 1.123  &  0.0869    &  8  & 0.69  & 0.038\\
\hline
\end{tabular}
\end{center}
\caption{Comparison between theory and numerical data for the circular geometry. Parameters $a_{2}$ and $a_{5}$ are determined with a non-linear least squares fit while parameters $a_{2}$,$a_{3}$,$a_{4}$, were fixed from theory predictions. The parameterization is given by equation \ref{sph-param}. $RMSR$ is the root mean squared residuals from the non-linear least squares fit.  The parameter $a_5$ appears to be relatively insensitive of the Renyi order, while $a_2$ does not. \label{sph-fit} }
\end{table}

\begin{figure}
\includegraphics[width=.45\textwidth]{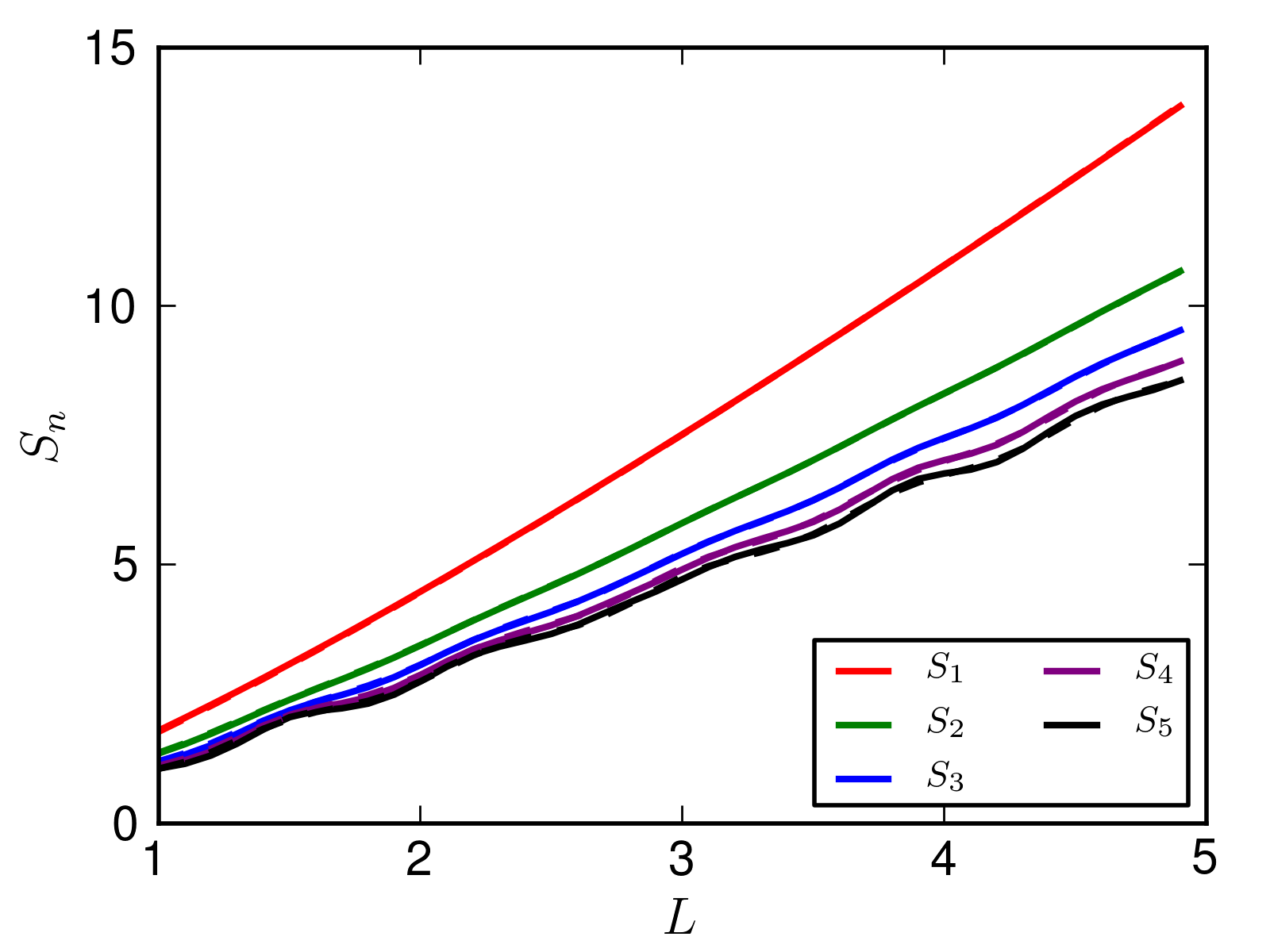}
\caption{Renyi entropies $S_1$ through $S_5$ for the 2D circular geometry. Solid lines are data and dashed lines are the fit whose parameters are found in Table \ref{sph-fit}. This fit lines fall on top of the data for low order $S_n$ and begin to deviate slightly for larger $S_n$. The amplitude and period of the oscillations, $a_3$ and $a_4$, are very accurately determined by the preceding analysis.}
\end{figure}



For the strip geometry we work with a slightly different density. We set our cell size to have length $W=\sqrt{N}$. Working in units where $k_f=2 \sqrt{\pi}$, we fit to the form,
\beq
S_n(W) = a_1 \ln{(W)} + a_2 + a_3 \frac{\cos{(a_4W+a_5)}}{W^{2/n+1/2}}.\label{cyl-param}
\eeq
Our analysis predicts
\beq
a_1=\frac{n+1}{3n}\frac{k_f L}{2\pi}
\eeq
(from the Widom fomula),
\beq
a_3=\frac{|f_n|L\sqrt{k_f}}{2\sqrt{\pi}(2k_f)^{2/n}}
\eeq
and $a_4=2k_f$.  We have chosen a fitting form for our data with two free parameters whose fits are provided in Table \ref{cyl-fit}. While the fits appear to be quite good there does appear to be small systematic differences between the theory and numerical data.  This is due to a slight underestimation of the amplitude of the oscillations which may be an artifact of the approximation to the integral over $\theta$ in equation \ref{theta_int_1} and \ref{theta_int_2}. However, the amplitude is qualitatively correct and the period is accurate. We expect that we can improve agreement between our analysis and numerics by increasing $W$, which is currently limited by computer memory. We note that direct comparisons between the quality of fit metrics for the two geometries and different order Renyi entropies need to be considered carefully as the scales of the entropies are not the same and the RMSR values are scale dependent.

\begin{table}
\begin{center}
\begin{tabular}{|c|c|c|c|c|c|c|}
\hline
$n $ & $a_1$ & $(a_2)_{fit}$ & $a_3$ &  $a_4$ & $(a_5)_{fit}$ & $RMSR$  \\ \hline
1 &10.48 &32.95 &  0        &  0     & 0      & 0.27\\
2 &7.86  &25.58 &  0.477    &  7.09  & 2.5 & 0.24\\
3 &6.99  &22.94 &  1.026    &  7.09  & 2.3 & 0.21\\
4 &6.55  &21.59 &  1.357    &  7.09  & 2.3 & 0.15\\
5 &6.29  &20.76 &  1.525    &  7.09  & 2.4 & 0.26\\
\hline
\end{tabular}
\end{center}
\caption{Comparison between theory and numerical data for the strip geometry.  Parameters $a_{2}$ and $a_{5}$ are determined with a NLLS fit while parameters $a_{2}$,$a_{3}$,$a_{4}$, were fixed from theory predictions.  The parameterization is given by equation \ref{cyl-param}.   Similar to the circular geometry, parameter $a_5$ appears to be independent of Renyi order while $a_2$ does not.\label{cyl-fit} }
\end{table}

\begin{figure}
\includegraphics[width=.45\textwidth]{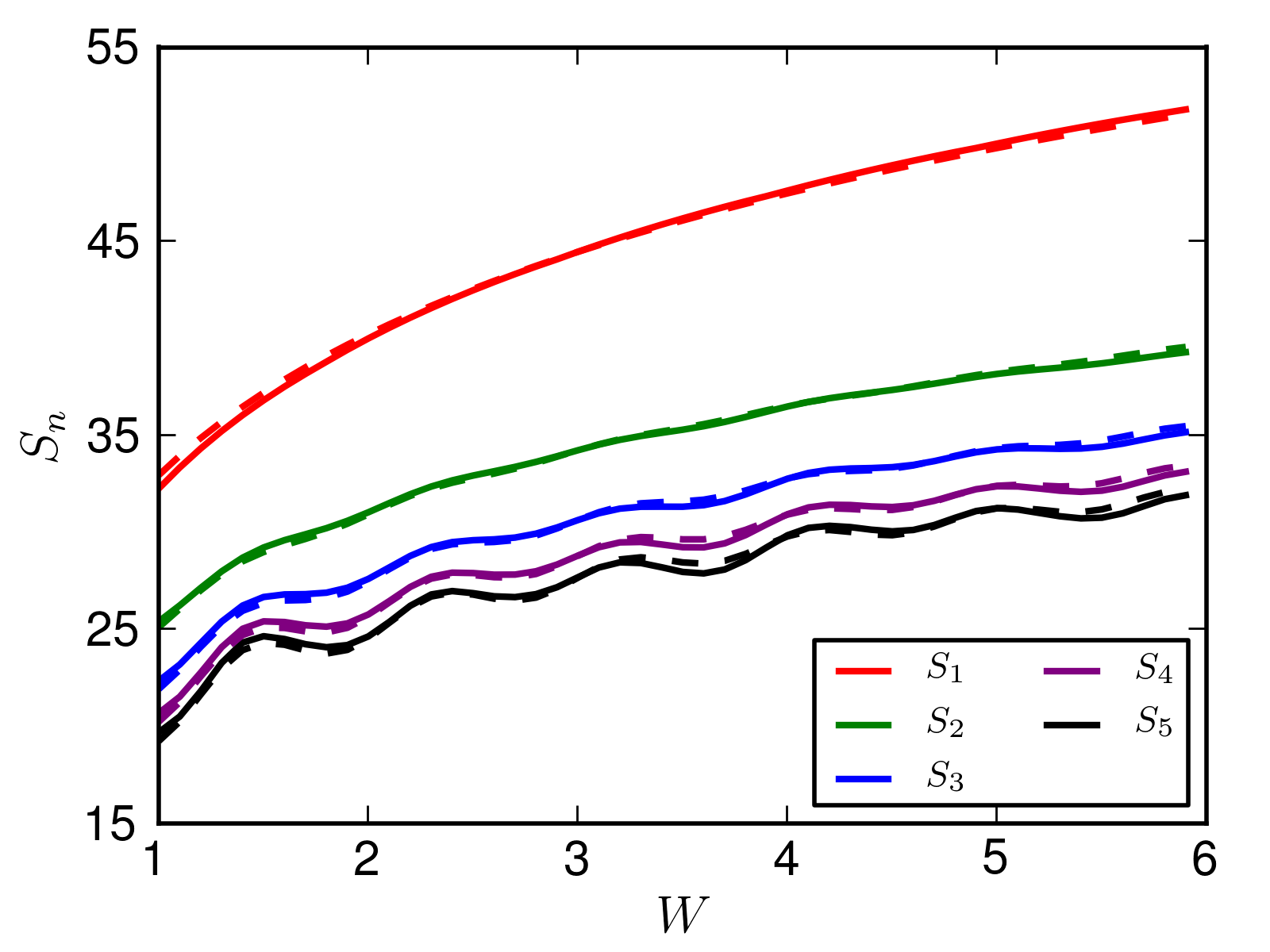}
\caption{Renyi entropies $S_1$ through $S_5$ for the 2D strip geometry. Solid lines are data and dashed lines are the fit parameterized in Table \ref{cyl-fit}. Here we note good agreement with the period, $a_4$, and acceptable agreement for the amplitude, $a_3$, with our analysis.  }
\end{figure}


Finally, let us mention what happens with interactions.  For this we may consider the data in Ref. \cite{qmc_renyi_FL}.  In that work Renyi entropies were computed for various interacting Fermi liquids as a function of interaction strength for both short-range interactions and Coulomb interaction using the swap operator in quantum Monte Carlo\cite{renyi_qmc_melko}.  For $S_2$, even at moderate interaction strengths (up to $r_s=1$ with Coulomb interactions),
 the free and interacting Renyi entropies agree quite well.  This data thus provides numerical evidence for our earlier theoretical claim that the period and power law prefactor of the decay is unchanged by weak to moderate interactions.
Additionally, as we described above, Ref. \cite{entsumrule} has shown that for a certain solvable model of an interacting Fermi
liquid, the proposed universality of Renyi entropies is indeed exactly true.

\section{Conclusions}

In this paper we have considered subleading oscillating terms in the Renyi entropy of free Fermi gases and Fermi liquids.  We give a simple analytic formula for the subleading oscillating term using the one-dimensional formulation of Fermi liquid entanglement.  This formula compares favorably with numerical calculations of the Renyi entropy of free Fermi gases in various geometries.  We also considered the effects of interactions and argued for a certain degree of universality in the subleading oscillating term.  Our arguments were checked by comparing to previous quantum Monte Carlo calculations of Renyi entropies in Fermi liquids as well as by comparing to exact results in a solvable model of a Fermi liquid.  Thus we have established precise agreement between theory and numerics for both the leading and subleading terms in the Renyi entropy of Fermi liquids.

An interesting future direction, which we have only just touched on here, is the exploration of the physics of the entanglement spectrum.  We know the full spectrum for a strip, but more generally it would be desirable to have an understanding the spectrum of more general regions and in the presence of interactions. Something like the bulk-edge correspondence for topological phases should be valid for Fermi liquids as well, but the precise form of this correspondence remains uncertain.  It would also be very interesting to extend our results to other kinds of quantum matter which support a Fermi surface but which may not be simple Fermi liquids.  Entanglement entropy in these models provide a conceptually simple probe of the Fermi surface, even if it is not associated with conventional electrons, and since both the oscillating term and the leading logarithmic term know about the Fermi surface geometry, we can extract the universal prefactor (analogous to the central charge) in front of the logarithmic term
in the Renyi entropy.  Finally, although we considered only spherical Fermi surfaces here, it is possible to extend our results to more general Fermi surface shapes.

\section{Acknowledgements}
JM and NMT were supported by the National Science Foundation under grant OCI-0904572.  BGS is supported by a Simons Fellowship through Harvard University.  This work used the Extreme Science and Engineering Discovery Environment (XSEDE), which is supported by National Science Foundation grant number OCI-1053575.

\newpage
\bibliography{osc_renyi_fl}

\appendix

\section{Computational Details \label{comp-detail}}
We compute the Renyi entropies from the eigenvalues of a spatially reduced density matrix using the correlation function technique \cite{CF-peschel}.
For a free particle Hamiltonian,
\beq
\hat H = -\sum_{m,n} \hat t_{m,n} c_n^\dagger c_m
\eeq
with $n$ and $m$ subsystem site indices, and $c_i$ and $c_i^\dagger$ the creation and annihilation operators for state $i$, eigenvalues for the density matrix can be computed using the relationship between it and the correlation function matrix. The correlation function matrix,
\beq
C_{ij}=\langle c_i^\dagger c_j\rangle,
\eeq
is determined entirely by the one particle operators. For this Hamiltonian we can write the density matrix as the exponential of a fictitious Hamiltonian, the entanglement Hamiltonian, with energy levels $\xi_k$ and single particle operators $a_k$ and $a_k^\dagger$,
\beq
\rho=\mathcal{K}\exp{\left(-\sum_k a_k^\dagger a_k\xi_k\right)}.
\eeq
where $\mathcal{K}$ is a normalization constant set by $\textrm{Tr} (\rho) = 1$. The new states, $a_k$ are related to the eigenvectors of the correlation matrix by, $c_i=\sum_k\phi_k(i)a_k$, and the eigenvalues of the entanglement Hamiltonian are related to those of the correlation matrix, $\zeta_i$, by,
\beq
\xi_i = \log\left(\frac{1-\zeta_i}{\zeta_i}\right).
\eeq
 The von Neumann Entropy, $S_1$, is then computed as,
\beq
  S_1 = \sum_i \ln\left(1 + \exp{(\xi_i)}\right) + \frac{-\xi_i}{1+\exp{(-\xi_i)}}
\eeq
and higher order Renyi entropies using a recursive formula,
\begin{equation}
\begin{array}{l}
 m_0=0\\
 w_0=1\\
 i=1\\
 \textrm{while i}<\textrm{N}_{tot}\\
 \quad w_i = w_{i-1}(1+\exp{(-\xi_i)})\\
 \quad m_i = m_{i-1}+\exp{(-n\xi_i)}(1+m_{i-1}) \\
 \quad i = i+1 \\
 S_n = \frac{1}{1-n} \log\left(\frac{m_{N_{tot}}+1}{(w_{N_{tot}})^n}\right)
\end{array}
\end{equation}
as shown in \cite{Renyi-mol}.

When computing high order Renyi entropies or the entropy of a large region, $m$ and $w$ can develop numerical instabilities. 
The eigenvalues of the correlation function matrix, $\zeta_i$, can be interpreted as the occupations of orbitals for region $A$ \cite{CF-peschel-2}.
 As the region size increases more of these orbitals become occupied with $\zeta_i\rightarrow1$ exponentially fast. The normalization factor, $w$, and the unnormalized trace, $m$, contain terms that diverge as the system or Renyi entropy order, $n$, grows large: $w=\prod_i (1-\zeta_i)^{-1}$ and $m\sim\prod_i (1-\zeta_i)^{-n}$.

While each term diverges individually, the ratio, $r_{m,w}=\frac{m_{N_{tot}}+1}{(w_{N_{tot}})^n}\rightarrow0$. This ratio can be written as a function of the Renyi entropies which has a known scaling form, $r_{m,w}=\exp{\left((1-n)S_n\right)}$. We can see that this ratio, $r_{m,w}$, scales as $W^{1-n}$ for cylindrical geometry and $L^{(1-n)L}$ for the spherical geometry, both of which go to zero as the system size increases. To deal with these numerical issues and maintain good accuracy for all system sizes, we use arbitrary precision libraries.


For the fits, very small $L$ and $W$ data are excluded and region sizes spanning less than one half the box size were used. This eliminates the contributions from region sizes which have very few particles on average and minimizes the interaction between periodic images of the subsystem. We fixed the parameters given by our analysis and fit the unknown parameters. The fitting procedure was performed using a non-linear least squares fit which minimized the sum of the residuals. 
We note that fits were also made without fixing the known parameters to their theoretical values, however the resulting values were determined to be too sensitive to the range of data included in the fit.

\end{document}